\documentclass[aps,prb,twocolumn,showpacs,amsmath,amssymb,superscriptaddress]{revtex4-1}

\usepackage{amsfonts}
\usepackage{amsmath}
\usepackage{graphicx}
\usepackage{bm}
\usepackage{amssymb}
\usepackage{lipsum}% for dummy text
\usepackage{times}
\usepackage{dcolumn}
\usepackage{cases}
\usepackage{braket}
\usepackage{booktabs}
\usepackage{setspace} 
\usepackage{color}

\usepackage{array}
\usepackage{float}
\usepackage{blindtext}
\newcommand{\singlefigwidthlarge}{0.48}

\makeatletter 
\renewcommand{\fnum@figure}{\textbf{Figure~\thefigure}}
\makeatother

\begin{document}

\title{Anisotropic Topological Hall Effect with Real and Momentum Space Berry Curvature in the Antiskrymion Hosting Heusler Compound Mn\textsubscript{1.4}PtSn}
\author{Praveen Vir}
\address{Max Planck Institute for Chemical Physics of Solids, D-01187 Dresden, Germany}
\author{Jacob Gayles}
\email[E-mail: ]{gayles@cpfs.mpg.de}
\address{Max Planck Institute for Chemical Physics of Solids, D-01187 Dresden, Germany}
\author{A. S. Sukhanov}
\address{Max Planck Institute for Chemical Physics of Solids, D-01187 Dresden, Germany}
\address{Institut f\"ur Festk\"orper- und Materialphysik, Technische Universit\"at Dresden, 01069 Dresden, Germany}
\author{Nitesh Kumar}
\address{Max Planck Institute for Chemical Physics of Solids, D-01187 Dresden, Germany}
\author{Fran\c{c}oise Damay}
\address{Laboratoire L\'eon Brillouin, CEA-CNRS, CEA Saclay, 91191 Gif-sur-Yvette, France}
\author{Yan Sun}
\address{Max Planck Institute for Chemical Physics of Solids, D-01187 Dresden, Germany}
\author{J\"urgen K\"ubler}
\address{Technische Universit\"at Darmstadt, Darmstadt, Germany}
\author{Chandra Shekhar}
\address{Max Planck Institute for Chemical Physics of Solids, D-01187 Dresden, Germany}
\author{Claudia Felser}
\email[E-mail: ]{Claudia.Felser@cpfs.mpg.de}
\address{Max Planck Institute for Chemical Physics of Solids, D-01187 Dresden, Germany}

\begin{abstract}
The topological Hall effect (THE) is one of the key signatures of topologically non-trivial magnetic spin textures, wherein electrons feel an additional transverse voltage to the applied current. The magnitude of THE is often small compared to the anomalous Hall effect. Here, we find a large THE of 0.9 $\mu\Omega$cm that is of the same order of the anomalous Hall effect in the single crystalline antiskyrmion hosting Heusler compound Mn\textsubscript{1.4}PtSn, a non-centrosymmetric tetragonal compound. The THE is highly anisotropic and survives in the whole temperature range where the spin structure is noncoplanar (\textless{}170 K). The THE is zero above the spin reorientation transition temperature of 170 K, where the magnetization will have a collinear and ferromagnetic alignment. The large value of the THE entails a significant contribution from the momentum space Berry curvature along with real space Berry curvature, which has never been observed earlier.
\end{abstract}

\maketitle

Spin-orbitronics combines microscopic control of magnetic textures with non-trivial Berry phase topology\cite{Manna2018,Xiao2010,Freimuth2013} and it is an emerging area for highly efficient next-generation devices\cite{Parkin2008,Fert2013}. The preliminary criteria for shortlisting of material is the strong spin-orbit coupling (SOC) and broken inversion symmetry that are the primary parameters to establish the Dzyaloshinskii-Moriya interaction (DMI) \cite{Moriya1960,Dzyaloshinsky1958}. The interplay between DMI with the symmetric exchange interactions, e.g. Heisenberg exchange, and an external magnetic field can stabilize a non-trivial topological magnetic texture with a finite real space Berry curvature\cite{Rossler2006,Taguchi2001}. A consequence of the real space Berry curvature\cite{Xiao2010,Nagaosa2010} is the topological Hall effect (THE) \cite{Taguchi2001,Bruno2004,Pfleiderer2010}. One class of magnetic textures that exhibit the THE are the so-called skyrmions, as has been observed firstly in the B20 compounds MnSi\cite{Neubauer2009} and FeGe\cite{Huang2012}. More recently, THE was also observed in the thin films of Heusler compounds Mn\textsubscript{2}RhSn\cite{Rana2016}, Mn\textsubscript{x}PtSn (x = 2 and 1.5) \cite{Li2018,Swekis2018}, and Mn\textsubscript{2}CoAl\cite{Ludbrook2017}. However, the presence of a skyrmion lattice is controversial as the thickness of the films approaches the bulk limit\cite{Rybakov2015}. In addition to skyrmion lattices, the THE can arise in noncoplanar spin structures\cite{Nagaosa2010} of the centrosymmetric compounds Mn\textsubscript{5}Si\textsubscript{3}\cite{Surgers2014} and Nd\textsubscript{2}Mo\textsubscript{2}O\textsubscript{7}\cite{Taguchi2001}. Interestingly, antiskyrmions have only recently been observed in the thin plates of tetragonal Heusler compounds Mn\textsubscript{1.4}PtSn and Mn\textsubscript{1.4}Pt\textsubscript{0.9}Pd\textsubscript{0.1}Sn with \emph{D}\textsubscript{2\emph{d}} symmetry\cite{nayak2017magnetic}.

Heusler compounds are known for their multi-functionality and tunability with controllable spin-orbit effects\cite{Manna2018,Wollmann2017} therefore, they are ideal materials for spin-orbitronic applications. The fundamental interactions such as SOC and exchange interaction can be microscopically controlled by chemical substitution\cite{Wollmann2017}. For example in Mn\textsubscript{x}YZ, the 3\emph{d} transition metal element, Mn, plays a crucial role in exchange interactions and the Y/Z elements are a source of large spin-orbit coupling. The compounds combination of Mn and Pt with other suitable group elements may lead to topological spin structures. It is therefore advantageous to utilize Heusler compounds to explore the underlying physics of the THE in real materials. The Heusler compound Mn\textsubscript{1.4}PtSn forms into a superstructure with \emph{D}\textsubscript{2\emph{d}} crystal symmetry\cite{Vir2018}, that allows for an anisotropic DMI\cite{Rossler2010} vector field and favors the stabilization of an antiskyrmion lattice in the \emph{ab} crystal plane of films\cite{Bogdanov2002}. In this article, we report a large and anisotropic THE in the single crystal of the superstructure Mn\textsubscript{1.4}PtSn, that arises from noncoplanar spin structure below 170 K. In the following, we show these exciting findings are well supported by neutron diffraction studies and first principles calculations. Interestingly, we demonstrate that the topological Hall effect has a significant contribution from the momentum space Berry curvature in addition to a more conventional real space Berry curvature for certain crystallographic directions.

\begin{figure}
 \includegraphics[width=\singlefigwidthlarge\textwidth]{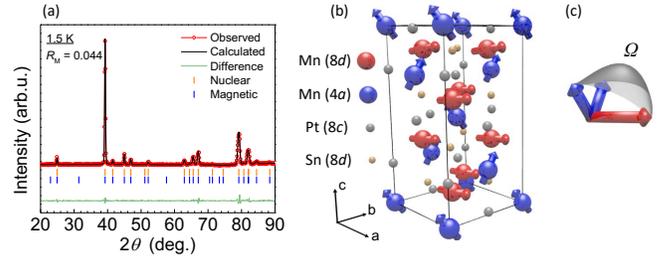}%
 \caption{(Color online) (a) Neutron diffraction pattern of Mn\textsubscript{1.4}PtSn measured at 1.5 K. (b) an exaggerated schematic of the noncoplanar magnetic structure, and (c) an exaggerated schematic of the noncoplanar spin arrangement of spins making a solid angle, $\it{\Omega}$, giving rise to emergent field.}
 \label{Fig1}
 \end{figure}

\begin{figure}
 \includegraphics[width=\singlefigwidthlarge\textwidth]{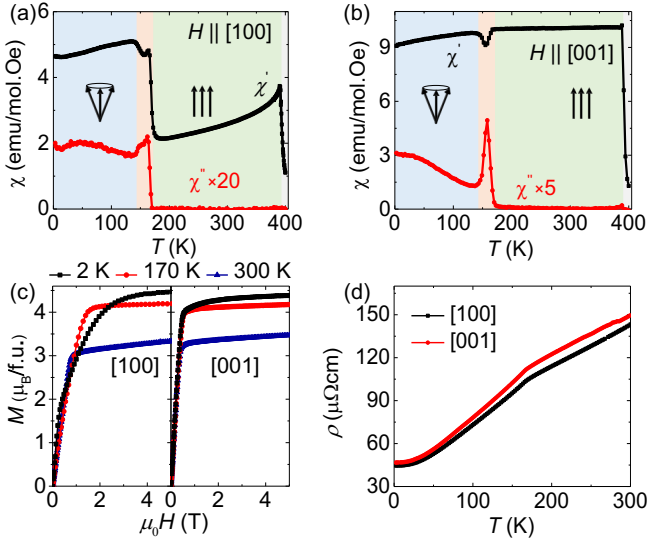}%
\caption{(Color online) Real (black) and imaginary (red) part of AC susceptibility measured along (a) {[}100{]} and (b) {[}001{]}. The shaded regions detail the magnetic structure, blue for noncoplanar, green for ferromagnetic, and the orange for a mixed state. (c) The isothermal magnetization curve for temperatures 2 K (black), 170 K (red) and 300 K (blue) (d) electrical resistivity measured for the directions of {[}100{]} (black) and {[}001{]} (red).}
 \label{Fig2}
 \end{figure}
 
 The Hall effects are insightful to obtain comprehensive information on the external field dependent magnetic texture in a given sample. In general, the measured Hall resistivity, \(\rho_{{yx}}\ \)of a topological magnetic material can be expressed by an empirical relation as, \(\rho_{{yx}} = \ \rho_{{yx\ }}^{\text{NHE}} + \rho_{{yx\ }}^{\text{AHE}} + \rho_{{yx\ }}^{\text{THE}}\text{.\ }\)Where, the normal Hall effect (NHE) is determined by the relation \(\rho_{{yx\ }}^{\text{NHE}} = \ R_{0}\mu_{0}H\), and is due to the Lorentz force in which \(R_{0}\) is the normal Hall coefficient and \(\mu_{0}H\) is the applied external magnetic field. The anomalous Hall effect (AHE) in a magnetic material can also be separated and is given by the relation\cite{Nagaosa2010}, \(\rho_{{yx\ }}^{\text{AHE}} = \left( \alpha\rho_{{yy}} + S_{A}\rho_{{yy}}^{2} \right)M\). Where $\alpha$ and \(S_{A}\) correspond to the extrinsic contribution from skew scattering and the impurity density independent contributions, respectively. \emph{M} is the magnetization and \(\rho_{{yy}}\) is the longitudinal resistivity. The intrinsic AHE originates from the momentum space Berry curvature, which is measured in the high field polarized regime\cite{Nagaosa2010}. Lastly, $\rho_{yx}^{\textrm{THE}}$ is the topological Hall resistivity which arises due to any noncoplanar order.
 
 Mn\textsubscript{1.4}PtSn is a noncentrosymmetric tetragonal compound (space group: \(I\overline{4}2d\)) with \emph{D}\textsubscript{2\emph{d}} symmetry. The lattice parameters \emph{a} and \emph{c} are 6.62 and 12.24 \AA, respectively. We investigate the magnetic structure with powder neutron diffraction at \emph{T} = 410 K (above \emph{T}$_C$), at \emph{T} = 200 K (above \emph{T}$_{\textrm SR}$), and at \emph{T} = 1.5 K. The pattern collected in the paramagnetic state contains only nuclear scattering and agrees with the crystal structure parameters inferred from X-ray diffraction measurements. Just above the spin reorientation, additional intensity peaks are observed at the positions of some nuclear reflections and the observed magnetic structure is characterized by a propagation vector k = (0, 0, 0). Below  \emph{T}$_{\textrm SR}$, significantly enhanced intensity of the (004) Bragg peak evidences the presence of spin canting. In Fig.~\ref{Fig1} (a), we show the powder neutron diffraction pattern (shown in red) of Mn\textsubscript{1.4}PtSn at 1.5 K. The refinement of this pattern provides two possible models of the magnetic structure; one that is noncollinear but coplanar and the other is a noncoplanar spin structure. Both the models give good agreement with the collected data. A coplanar spin structure gives rise to a zero scalar spin chirality and therefore cannot exhibit finite THE. With first principle methods we have relaxed the magnetic structure for Mn$_{1.5}$PtSn which shows a noncoplanar groundstate, with the Mn ions at the 8{\it d} Wyckoff position align collinearly in the {\it ab}-plane while the Mn atoms at the 4{\it a} position order antiferromagnetically with an angle of $\sim$14$^\circ$.  The observation of THE in combination with the theoretical calculations confirms the existence of a noncoplanar spin structure. Based on the neutron diffraction pattern (see supporting information), we show the magnetic unit cell of Mn\textsubscript{1.4}PtSn in Fig.~\ref{Fig1} (b). In the crystal structure, Mn atoms reside at two Wyckoff positions 8\emph{d} (in red) and 4\emph{a} (in blue), forming three magnetic sublattices. At the fully occupied site 8\emph{d,} the spin of Mn atoms arranges collinearly in the \emph{ab}-plane, while at the partially (88 percent) resided position 4\emph{a}, the Mn atoms are antiferromagnetic and noncoplanar to the 8\emph{d} site. The nonmagnetic atoms Pt and Sn occupy the 8\emph{c} and 8\emph{d} site, respectively. A schematic view of the noncoplanar spin structure with $\it{\Omega}$ is shown in Fig.~\ref{Fig1} (c).

Fig.~\ref{Fig2} (a) and (b) show the AC susceptibility (both the real ($\chi'$) and imaginary ($\chi''$) parts) as a function of temperature for the directions \emph{H} \textbar{}\textbar{} {[}100{]} and \emph{H} \textbar{}\textbar{} {[}001{]}, respectively. We observe two distinct changes in $\chi'$ (black curve) at 392 K and 170 K independent of the direction of the applied magnetic field. The sharp change at high temperature corresponds to the transition from a paramagnetic to a collinear ferromagnetic state (green background). The change in $\chi'$ at \emph{T} = 170 K corresponds to the spin-reorientation transition temperature ($T_{\textrm{SR}}$). Below the $T_{\textrm{SR}}$ (blue background), the spins cant and stabilize into a noncoplanar spin structure. A dip in the $\chi'$ is also observed in the vicinity of $T_{\textrm{SR}}$ (yellow background) extending from 170 K to 135 K. This intermediate regime is a combination of the collinear ferromagnet and the noncoplanar spin structure. The neutron diffraction studies show that the canting in this regime is smaller than that in the low-temperature regime. For \emph{H} \textbar{}\textbar{} {[}100{]}, $\chi'$ exhibits a sudden increase when the temperature decreases below $T_{\textrm{SR}}$, which indicates the emergence of an in-plane magnetization due to canting. In contrast, the value of $\chi'$ is nearly constant for \emph{H} \textbar{}\textbar{} {[}001{]} across $T_{\textrm{SR}}$. For both the field directions, the imaginary part of the susceptibility ($\chi''$) (red curves) also displays a spin reorientation transition. In Fig.~\ref{Fig2} (c), we show the isothermal magnetization at 2 K, 170 K, and 300 K for both \emph{H} \textbar{}\textbar{} {[}100{]} and \emph{H} \textbar{}\textbar{} {[}001{]}. The saturation magnetization is found to be 4.7 $\mu$\textsubscript{B}/f.u. at 2 K. The large net moment is beneficial for the extraction of the relevant transport quantities as compared to those of the skyrmionic compounds MnSi and FeGe, with moments that range between 0.3 and 1.0 $\mu$\textsubscript{B}/f.u.\cite{Neubauer2009,Huang2012}. This is further supported by our first principle calculations, which show a local Mn exchange splitting of \(\approx\)4 eV in Mn\textsubscript{1.4}PtSn, (see supporting information) compared to \(\approx\)1 eV in Mn\textsubscript{1-x}Fe\textsubscript{x}Ge\cite{Gayles2015}. In the temperature range where the spin structure is noncollinear, the increase in magnetization is nonlinear with the magnetic field below saturation. Above $T_{\textrm{SR}}$ the variation of the moment is linear in the magnetic field. The easy axis of the magnetization is along the {[}001{]} direction, above and below $T_{\textrm{SR}}$. Our first principles calculations show the magnetic anisotropy between the \emph{ab}-plane and the \emph{c}-axis favors the \emph{c}-axis by 4 meV/u.c. In Fig.~\ref{Fig2} (d) we show the resistivity as a function of temperature for the applied current along {[}100{]} (black curve) and {[}001{]} (red curve). The compound behaves as a metal with a slightly higher resistivity along the \emph{c}-axis compared to the \emph{ab}-plane.

 \begin{figure}[t!]
\includegraphics[width=\singlefigwidthlarge\textwidth]{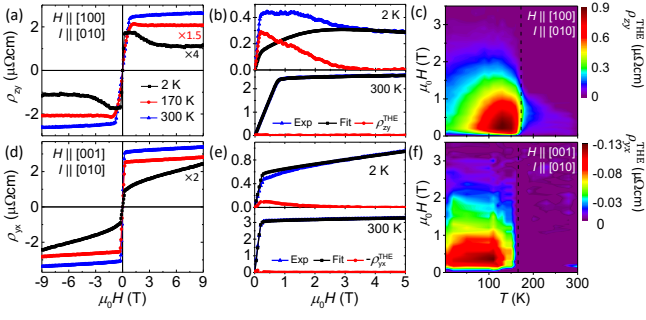}
\caption{(Color online) The top and bottom rows show the Hall transport for {\it H} \textbar{}\textbar{} {[}100{]} and  {\it H} \textbar{}\textbar{} {[}001{]}, respectively. In (a) and (d) the measured Hall resistivity a temperatures 2 K (black), 170 K (red), and 300 K (blue). In (b) and (e) the measured Hall resistivity (blue), the high field extrapolated AHE+NHE (black) and  the subtraction of the two resulting in THE (red) for temperature 2 and 300 K in the top and bottom panels, respectively. In (c) and (f) contour plots of THE for temperature vs. field.}
\label{Fig3}
\end{figure}

In order to investigate the topological aspects of the noncoplanar spin structure in Mn\textsubscript{1.4}PtSn, we performed Hall resistivity and longitudinal resistivity measurements. Fig.~\ref{Fig3} (a) and (d) show the field dependent Hall resistivity at the temperatures 2, 170 and 300 K, for \emph{H} \textbar{}\textbar{} {[}100{]}, \emph{I} \textbar{}\textbar{} {[}010{]} and \emph{H} \textbar{}\textbar{} {[}001{]}, \emph{I} \textbar{}\textbar{} {[}010{]}. Below $T_{\textrm{SR}}$ (2 K, black curve and 170 K, red curve), there is a distinct magnetization-independent increase of the Hall effect separate from the normal and anomalous Hall effects, which is a clear indication of THE in this compound. The magnitude of the THE depends on temperature and appears only below $T_{\textrm{SR}}$ (blue and red curves) while the anomalous Hall effect is observed up to \emph{T}\textsubscript{C}. The THE is comparatively larger for \emph{H} \textbar{}\textbar{} {[}100{]} than for \emph{H} \textbar{}\textbar{} {[}001{]}. The longitudinal resistivity \(\rho_{{yy}}\) decreases with the applied magnetic field giving rise to negative magnetoresistance for both \emph{H} \textbar{}\textbar{} {[}100{]} and \emph{H} \textbar{}\textbar{} {[}001{]} orientations (see Fig. S2). In order to get a clearer picture of the THE in Mn\textsubscript{1.4}PtSn we extract the topological contribution from the overall Hall resistivity. In Fig.~\ref{Fig3} (b) the upper and lower panels show the measured \(\rho_{{zy}}\) (blue) at 2 K and 300 K, respectively for the configuration \emph{H} \textbar{}\textbar{} {[}100{]}; \emph{I} \textbar{}\textbar{} {[}010{]}. The contribution of the NHE (\(R_{0}\mu_{0}H\)) + AHE (\(S_{A}M\rho_{{yy}}^{2}\)) (shown as black curves) in the high field regime is subtracted from the measured $\rho_{zy}$. Thus, the remaining value is THE (red curve) with a maximum of 0.3 $\mu\Omega$cm at 2 K. The same analysis was followed for 300 K ($T \textgreater{}T_{\textrm{SR}}$) and it can be seen in the lower panel of Fig.~\ref{Fig3} (b) that the measured \(\rho_{{zy}}\) matches well with the fitted NHE+AHE, thus resulting in a zero THE contribution.  Fig.~\ref{Fig3} (c), represents contour plot for THE in which THE reaches the maximum value of 0.9 $\mu$â$\Omega$cm at a field of 0.3 T and a temperature of 130 K. The corresponding analysis for \emph{H} \textbar{}\textbar{} {[}001{]} is shown in  Fig.~\ref{Fig3} (e) and the resulting values of THE are plotted as a contour in  Fig.~\ref{Fig3} (f). The THE for \emph{H} \textbar{}\textbar{} {[}001{]} is significantly smaller and has an opposite sign compared to that of \emph{H} \textbar{}\textbar{} {[}100{]}. The topological Hall resistivity is positive and negative for \emph{H} \textbar{}\textbar{} {[}001{]} and \emph{H} \textbar{}\textbar{} {[}100{]}, respectively.

\begin{figure}[t!]
\includegraphics[width=\singlefigwidthlarge\textwidth]{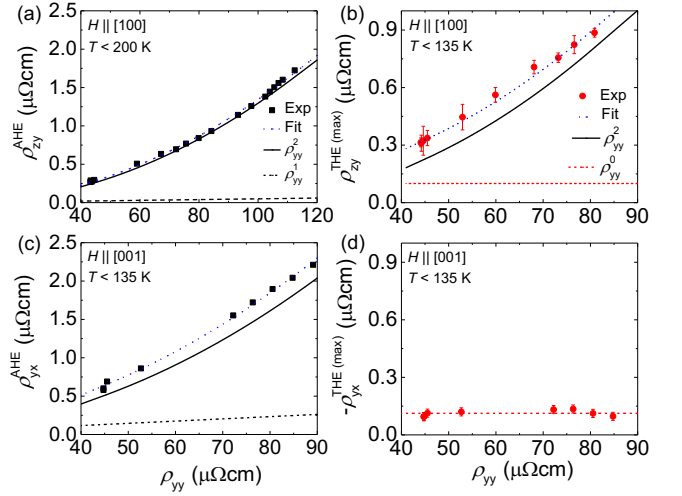}
\caption{(Color online) (a) and (c) The anomalous Hall resistivity as a function of the longitudinal resistivity, \(\rho\) (black squares). The fitting in terms of (dashed black) \(\rho_{yy}^{1}\ \)and \(\rho_{yy}^{2}\ \)(solid black), and the summation (dashed blue). (b) and (d) The topological Hall resistivity as a function of the (red circles). The fitting in terms of (dashed red) \(\rho_{yy}^{0}\ \)and \(\rho_{yy}^{2}\ \)(solid red), and the summation (dashed blue).}
\label{Fig4}
\end{figure}

In Fig.~\ref{Fig4} we present the AHE and the THE as a function of the longitudinal resistivity for \emph{H} \textbar{}\textbar{} {[}100{]} and \emph{H} \textbar{}\textbar{} {[}001{]}. We use a power law of Equation (1), to deconvolute the various components of AHE. In the case of \emph{H} \textbar{}\textbar{} {[}100{]}, \(\rho_{{zy}}^{\text{AHE\ }}\) varies as a square function of \(\rho_{{yy}}\) with a negligible contribution from the linear term (see  Fig.~\ref{Fig4} (a)). This indicates that the AHE is determined by the intrinsic Berry curvature component. This is further validated by a fairly good agreement between the experimentally measured and theoretically calculated values of the anomalous Hall conductivity (AHC) as will be discussed in the next section. For \emph{H} \textbar{}\textbar{} {[}001{]},\(\ \rho_{{yx}}^{\text{AHE\ }}\) also shows a strong intrinsic contribution, however, there is a significant contribution that is linear in \(\rho_{{yy}}\ \)(see  Fig.~\ref{Fig4} (c)). This indicates a skew scattering contribution to the AHE, which is consistent with the enhanced scattering in this direction indicated by a larger resistivity value (Figure 2(d)).

In contrast to the NHE and AHE, the topological Hall resistivity is maximized in the low external field limit. Due to this, the THE can be quite cumbersome to extract from the total Hall resistivity, and even more so to distinguish between momentum and real space contributions\cite{Hoffmann2015,Kubler2014}. The momentum space and real space contributions can be separated into two regimes where the SOC does not significantly influence the scattering lifetime of the electron mean free path\cite{Onoda2004,Nagaosa2012}. In the case of a magnetic texture on the order of the periodic crystal structure and much smaller than the mean free path of the electron, the momentum space Berry curvature dominates, where the role of SOC is replaced by the variation of the noncoplanar magnetic texture in small external fields\cite{Nagaosa2006}. This emergent field is the result of the cone angle subtended by three spins that give rise to the scalar spin chirality, given as $\chi_{ij}=\sum\widehat{S}_i\cdot(\widehat{S}_j\times\widehat{S}_k)$, with $\widehat{S}_i$ as the spin vector of the magnetic atom. The size and direction of the emergent field are determined by the external magnetic field for the noncoplanar systems and is approximately linear in the variation of the external magnetic field at a given temperature.

In the second case of an adiabatic topological magnetic structure where the mean free path is much smaller than the length scale of the magnetic texture, the THE is due to the continuum limit ($\widehat{S}_i~\rightarrow~\widehat{n}$) of the scalar spin chirality or the so-called topological winding number\cite{Bruno2004,Neubauer2009}. $\widehat{n}$ is the local unit vector direction of the magnetization. Here, the \(\rho_{{yx\ }}^{\text{THE}}\ \)can be expressed as\(\ R_{{yx}}^{\text{THE}}B_{\text{em}}\), where \(R_{{yx}}^{\text{THE}}\) is the topological Hall coefficient and \(B_{\text{em}}\) is the emergent magnetic field due to real-space Berry curvature\cite{Freimuth2013,Franz2014}. This field adiabatically deflects the spin-oriented motion of the electrons with opposite sign of the field for the respective spins\cite{Franz2014}. In the case of a skyrmionic textures the emergent field is constant and for an external field range.

\begin{figure}[th]
\includegraphics[width=\singlefigwidthlarge\textwidth]{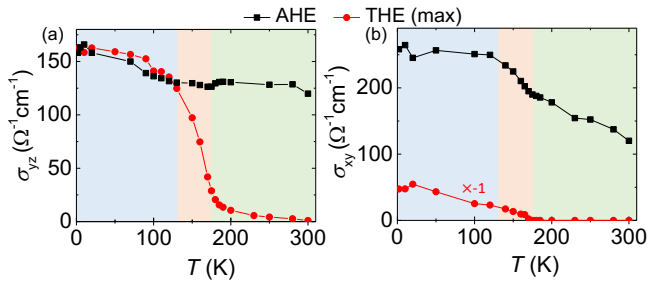}
\caption{(Color online) Anomalous and topological Hall conductivity for collinear (green) and noncoplanar (light blue, yellow) magnetic regions for configuration (a) {\it H} \textbar{}\textbar{} {[}100{]} and (b) {\it H} \textbar{}\textbar{} {[}001{]}.}
\label{Fig5}
\end{figure}

 Fig.~\ref{Fig4} (b) and (d) show the experimentally measured maximum value of the topological Hall resistivity (red circles) in the temperature range of 2--135 K. The THE when plotted in terms of $\rho_{yy}$ is anisotropic between the two field directions, even more so than the AHE. In  Fig.~\ref{Fig4} (d), \(\rho_{{yx}}^{\text{THE\ }(\max)}\) is constant and independent of $\rho_{yy}$ for \emph{H} \textbar{}\textbar{} {[}001{]}. This is consistent with the previous reports, which show that the real space Berry curvature induced-THE is scattering independent and therefore, is constant with longitudinal resistivity\cite{Tatara2007}. The $\rho_{yy}$-independent THE component comes from \(R^{\text{THE}}B_{\text{em}}\). This behavior points to a large-scale topological structure such as the expected antiskyrmion, due to the DMI in the {\it ab}-plane that contributes in the formation of long range spin textures. For \emph{H} \textbar{}\textbar{} {[}100{]}, in addition to $\rho_{yy}$-independent contribution, the dominant contribution to \(\rho_{{zy}}^{THE\ (max)}\), varies as a square of $\rho_{yy}$. The contribution that relates \(\rho_{{yy}}^{2}\) is independent of the net magnetization, analogous to the intrinsic Berry curvature contribution of AHE that is dependent on the magnetization. This \(\rho_{{yy}}^{2}\) contribution is due to the finite spin chirality product. In \emph{H} \textbar{}\textbar{} {[}100{]} there is an unexpected $\rho_{yy}$-independent term, that may arise from dipole-dipole interactions that may also stabilize long range spin textures. Furthermore,The $\rho_{yy}$-independent term for \emph{H} \textbar{}\textbar{} {[}001{]} and \emph{H} \textbar{}\textbar{} {[}100{]} has approximately the same magnitude with opposite sign. Recently, in thin films of Mn$_{1.5}$PtSn the THE was observed\cite{Swekis2018}, however the fundamental origin of the THE, the anisotropy in the transport, and the details of the magnetic structure were not fully understood due to the limitations of the thin film set up.
 
For \(R^{\text{THE}}\) we calculate the difference of the spin-resolved Hall conductivity tensors in the semiclassical limit for a collinear ferromagnet\cite{Franz2014}, which provides further evidence that \(R^{\text{THE}}\) changes sign with the crystallographic direction (supporting information). The calculated values of \(R^{\text{THE}}\) are -1.5 and 0.5 (â$\Omega$cmT\textsuperscript{-1})$\times$10\textsuperscript{-8} for \emph{H} \textbar{}\textbar{} {[}001{]} and \emph{H} \textbar{}\textbar{} {[}100{]}, respectively. This would correspond to \(B_{\text{em}} \approx 10\ T\), to match the experimental topological Hall resistivities. Lastly, in the vicinity of the $T_{\textrm{SR}}$, the THE scales linearly with \(\rho_{{yy}}\) (supporting information). This can be attributed to a term analogous to the skew scattering contribution in AHE\cite{Nagaosa2010,Ishizuka2018}. Above $T_{\textrm{SR}}$, the THE monotonically decreases to zero for \emph{H} \textbar{}\textbar{} {[}100{]}. However, in case of \emph{H} \textbar{}\textbar{} {[}001{]} the THE shows to be strictly zero above the $T_{\textrm{SR}}$ where antiskyrmions are suspected to be stable\cite{nayak2017magnetic}. This strict requirement of zero THE is possibly due to the symmetry of the antiskyrmion in combination with the electronic structure. It is noteworthy that a combination of the momentum space and real space Berry curvature induced THE has not been identified in a single material and allows for the possibility to tune the size of these effect by the electronic structure.

The anomalous Hall conductivity
(AHC),\(\ \sigma_{{xy}}^{\text{AHE}}\), can be extracted from the
measured Hall resistivity, the anomalous Hall resistivity, and the
longitudinal resistivity, using the formula,

\begin{equation}
\sigma_{{xy}}^{\text{AHE}} = \ \frac{\rho_{{yx}}^{\text{AHE}}}{\rho_{{yx}}^{2} + \ \rho_{{yy}}^{2}}.\\
\label{ahc}
\end{equation}

The topological Hall conductivity (THC) can analogously be extracted using the same method as Eq.~\ref{ahc}. In Fig.~\ref{Fig5} (a) and (b) the temperature dependent AHC (black curve) and THC (red curve) is shown for both \emph{H} \textbar{}\textbar{} {[}100{]} and \emph{H} \textbar{}\textbar{} {[}001{]}. We find the highest value of AHC 165 and 250 â$\Omega$\textsuperscript{-1}cm\textsuperscript{-1} at 5 K for \emph{H} \textbar{}\textbar{} {[}100{]} and \emph{H} \textbar{}\textbar{} {[}001{]}, respectively. In our first principle calculations with a Wannier interpolated Hamiltonian, we use the Kubo formula to calculate AHC values of 155 and 232 â$\Omega$\textsuperscript{-1}cm\textsuperscript{-1}, for the respective collinear magnetization directions. Here is it clearly seen that the Berry curvature is anisotropic in momentum space, as is expected for a tetragonal magnetic system. Furthermore, the THC is of the same order of AHC for \emph{H} \textbar{}\textbar{} {[}100{]} and indicates the main contribution is intrinsic for temperatures below the $T_{\textrm{SR}}$. Whereas, the THC for \emph{H} \textbar{}\textbar{} {[}001{]} is strongly dependent on the temperature and linearly decreases to zero at the boundary of $T_{\textrm{SR}}$ region.

We observed a large THE in a noncoplanar structure for the field in the \emph{ab}-plane. The THE is of the same order of magnitude as the AHE and clearly visible in our experiments. We show that the THE is a function of the square of longitudinal resistivity which provides a new direction of understanding of THE in a magnetic material. The value of the THE in these bulk compounds can be tuned by the temperature, going from a large value to zero below and above the $T_{\textrm{SR}}$, respectively in addition to a change in sign and magnitude for the field direction. The anisotropic behavior of the THE implies that the electronic structure also plays a key role for the scattering of spin-dependent electrons at the Fermi surface. This leads to a pathway to engineering large THE in Heusler compounds exhibiting antiskyrmion-type spin structures. Therefore, it is plausible to additionally, tune the magnetic textures and magnitude of these effects by tuning the electronic structure and spin-orbit coupling by chemical doping.

This work was financially supported by the ERC Advanced Grant 742068 ``TOPMAT''. We acknowledge funding by the Deutsche Forschungsgemeinschaft (DFG, German Research Foundation) under SPP 2137 (Project number 403502666). We thank Vivek Kumar, Peter Adler and Manfred Reehuis for fruitful discussion on refinement of neutron diffraction of Mn\textsubscript{1.4}PtSn.

\end{document}